\documentstyle[12pt]{article}
\catcode`\@=11

\def\@normalsize{\@setsize\normalsize{15pt}\xiipt\@xiipt
\abovedisplayskip 14pt plus3pt minus3pt%
\belowdisplayskip \abovedisplayskip
\abovedisplayshortskip  \z@ plus3pt%
\belowdisplayshortskip  7pt plus3.5pt minus0pt}
\def\small{\@setsize\small{13.6pt}\xipt\@xipt
\abovedisplayskip 13pt plus3pt minus3pt%
\belowdisplayskip \abovedisplayskip
\abovedisplayshortskip  \z@ plus3pt%
\belowdisplayshortskip  7pt plus3.5pt minus0pt
\def\@listi{\parsep 4.5pt plus 2pt minus 1pt
            \itemsep \parsep
            \topsep 9pt plus 3pt minus 3pt}}

\def\underline#1{\relax\ifmmode\@@underline#1\else
        $\@@underline{\hbox{#1}}$\relax\fi}
\@twosidetrue
\relax

\catcode`@=12

\catcode`\@=11

\def\section{\@startsection{section}{1}{\z@}{3.5ex plus 1ex minus
   .2ex}{2.3ex plus .2ex}{\large\bf}}

\def\ps@headings{\def\@oddfoot{}\def\@evenfoot{}
\def\@oddhead{\hbox{}\hfill
        \makebox[.5\textwidth]{\raggedright\ignorespaces --\thepage{}--
        \hfill }}
\def\@evenhead{\@oddhead}
\def\subsectionmark##1{\markboth{##1}{}}
}

\ps@headings

\catcode`\@=12

\relax

\def\figcap{\section*{Figure Captions\markboth
        {FIGURECAPTIONS}{FIGURECAPTIONS}}\list
        {Fig. \arabic{enumi}:\hfill}{\settowidth\labelwidth{Fig. 999:}
        \leftmargin\labelwidth
        \advance\leftmargin\labelsep\usecounter{enumi}}}
 \relax
\def\tablecap{\section*{Table Captions\markboth
        {TABLECAPTIONS}{TABLECAPTIONS}}\list
        {Table \arabic{enumi}:\hfill}{\settowidth\labelwidth{Table 999:}
        \leftmargin\labelwidth
        \advance\leftmargin\labelsep\usecounter{enumi}}}
 \relax
\def\reflist{\section*{References\markboth
        {REFLIST}{REFLIST}}\list
        {[\arabic{enumi}]\hfill}{\settowidth\labelwidth{[999]}
        \leftmargin\labelwidth
        \advance\leftmargin\labelsep\usecounter{enumi}}}
 \relax

\catcode`\@=11

\def\marginnote#1{}
\newcount\hour
\newcount\minute
\newtoks\amorpm
\hour=\time\divide\hour by60
\minute=\time{\multiply\hour by60 \global\advance\minute by-
\hour}
\edef\standardtime{{\ifnum\hour<12 \global\amorpm={am}%
    \else\global\amorpm={pm}\advance\hour by-12 \fi
    \ifnum\hour=0 \hour=12 \fi
    \number\hour:\ifnum\minute<100\fi\number\minute\the\amorpm}}
\edef\militarytime{\number\hour:\ifnum\minute<100\fi\number\minute}
\def\draftlabel#1{{\@bsphack\if@filesw {\let\thepage\relax
  \xdef\@gtempa{\write\@auxout{\string
    \newlabel{#1}{{\@currentlabel}{\thepage}}}}}\@gtempa
    \if@nobreak \ifvmode\nobreak\fi\fi\fi\@esphack}
     \gdef\@eqnlabel{#1}}
\def\@eqnlabel{}
\def\@vacuum{}
\def\draftmarginnote#1{\marginpar{\raggedright\scriptsize\tt#1}}
\def\draft{\oddsidemargin -.5truein
        \def\@oddfoot{\sl preliminary draft \hfil
        \rm\thepage\hfil\sl\today\quad\militarytime}
        \let\@evenfoot\@oddfoot \overfullrule 3pt
        \let\label=\draftlabel
        \let\marginnote=\draftmarginnote
   
\def\@eqnnum{(\theequation)\rlap{\kern\marginparsep\tt\@eqnlabel}%
\global\let\@eqnlabel\@vacuum}  }
\def\preprint{\twocolumn\sloppy\flushbottom\parindent 1em
        \leftmargini 2em\leftmarginv .5em\leftmarginvi .5em
        \oddsidemargin -.5in    \evensidemargin -.5in
        \columnsep 15mm \footheight 0pt
        \textwidth 250mmin      \topmargin  -.4in
        \headheight 12pt \topskip .4in
        \textheight 175mm
        \footskip 0pt
        
\def\@oddhead{\thepage\hfil\addtocounter{page}{1}\thepage}
        \let\@evenhead\@oddhead \def\@oddfoot{} \def\@evenfoot{} 
}
\def\titlepage{\@restonecolfalse\if@twocolumn\@restonecoltrue\onecolumn
     \else \newpage \fi \thispagestyle{empty}\c@page\z@
        \def\thefootnote{\fnsymbol{footnote}} }
\def\endtitlepage{\if@restonecol\twocolumn \else  \fi
        \def\thefootnote{\arabic{footnote}}
        \setcounter{footnote}{0}}  %\c@footnote\z@ }
\catcode`@=12
\relax
\def\ps@headings{\def\@oddfoot{}\def\@evenfoot{}
\def\@oddhead{\hbox{}\hfill
        \makebox[.5\textwidth]{\raggedright\ignorespaces --\thepage{}--
        \hfill }}
\def\@evenhead{\@oddhead}
\def\subsectionmark##1{\markboth{##1}{}}
}

\ps@headings

\relax

\def\firstpage#1#2#3#4#5#6{
\begin{document}
%%%%%%%%%%%%%%%%%%%%%%%%%%%%%%%%%%%%%%%%
%\draft
%%%%%%%%%%%%%%%%%%%%%%%%%%%%%%%%%%%%%%%%
\begin{titlepage}
\nopagebreak
\title{\begin{flushright}
        \vspace*{-1.8in}
        {\normalsize CERN-TH/98-46}\\[-9mm]
%       {\normalsize DEMO-HEP.98.02--#1 -- #2\\[-9mm]
   {\normalsize DEMO-HEP.98-02}\\[-9mm]
        {\normalsize hep-th/9802107}\\[4mm]
\end{flushright}
%\vfill
\vspace{3cm}
{#3}}
\author{\large #4 \\[0.0cm] #5}
\maketitle
%\vskip -7mm     
\vskip 5mm
\nopagebreak 
\begin{abstract}
{\noindent #6}
\end{abstract}
\vfill
%\vspace{4cm}
\begin{flushleft}
\rule{16.1cm}{0.2mm}\\[-3mm]
%$^{\star}${\small Research supported in part by\vspace{0mm}
% the EEC under TMR contract ERBFMRX-CT96-0090.
%\\[-3mm]
$^\ast$e-mail: Alexandros.Kehagias@cern.ch \\[-3mm]
\hspace{1.45cm}{manolis@timaios.nrcps.ariadne-t.gr} \\[0mm]
CERN-TH-98-46\\[-1mm]
February 1998
\end{flushleft}
\thispagestyle{empty}
\end{titlepage}}

\def\simlt{\stackrel{<}{{}_\sim}}
\def\simgt{\stackrel{>}{{}_\sim}}
\newcommand{\dal}{\raisebox{0.085cm}
{\fbox{\rule{0cm}{0.07cm}\,}}}

\newcommand{\be}{\begin{eqnarray}}
\newcommand{\ee}{\end{eqnarray}}
\newcommand{\btau}{\bar{\tau}}
\newcommand{\p}{\partial}
\newcommand{\bp}{\bar{\partial}}
\renewcommand{\a}{\alpha}
\renewcommand{\b}{\beta}
\newcommand{\g}{\gamma}
\renewcommand{\d}{\delta}
\newcommand{\gsi}{\,\raisebox{-0.13cm}{$\stackrel{\textstyle
>}{\textstyle\sim}$}\,}
\newcommand{\lsi}{\,\raisebox{-0.13cm}{$\stackrel{\textstyle
<}{\textstyle\sim}$}\,}
\date{}
\firstpage{3118}{IC/95/34}
{\large 
{\Large E}IGHT-{\Large D}IMENSIONAL  {\Large S}ELF-{\Large D}UAL 
{\Large S}PACES$^\star$\\ 
%{\Large A}CTION 
%\\
%FOR THE {\Large T}YPE {\Large IIB} {\Large S}UPERSTRING$^\star$ \\
\phantom{X}}
{E.G. Floratos$^{a,b}$
 and  A. Kehagias${^{c}}^\ast$
\phantom{X} }
{\vspace{-.4cm}
\normalsize\sl $^a$
Institute of Nuclear Physics, NRCS Democritos, Athens, Greece\\
\vspace{-.4cm}
\normalsize\sl 
$^b$Physics Department, University of Iraklion, Crete, Greece. \\
\normalsize\sl $^c$ Theory Division, CERN, 1211 Geneva 23, Switzerland
}
{We discuss  higher-dimensional gravitational instantons by studying
appropriate self-duality equations for the spin connection.  
In seven and in eight dimensions, the corresponding  
spaces admit a covariantly constant spinor and, consequently, have holonomies 
in $G_2$ and $Spin(7)$, respectively. We find a non-compact 
solution to the  self-duality equations in eight dimensions in which the 
self-dual  space has  an elliptically-fibered structure.}
\newpage
\section{Introduction}

Four-dimensional gravitational instantons \cite{H}
have been constructed long-ago
and like  their  Yang-Mills ancestors, they are of finite action and 
provide tunnelling amplitudes for 
distinct gravitational vacua \cite{EGH}. 
In the supergravity context, gravitational 
instantons break half of the supersymmetries. For example, for a self-dual 
background in the $N=2$ theory, 
only the left-handed supersymmetry survives. The broken 
one generates fermionic zero modes and give mass to the right-handed 
gravitino. Of course, none of these backgrounds is asymptotically Euclidean 
 as follows from the positive action conjecture \cite{GHP}. 
According to the latter,
the action $S$ for any asymptotically Euclidean   space with everywhere
vanishing Ricci scalar satisfies $S\geq 0$ where the bound is saturated 
for flat space. However, this does not exclude the existence of 
asymptotically locally Euclidean (ALE) spaces. Making use of the latter 
possibility, instanton solutions have been constructed \cite{H},\cite{EH}. 

Yang-Mills instantons exist in higher dimensions as well 
\cite{CDFN}--\cite{GN}. 
For example, instantons in seven dimensions with gauge group $G_2$ \cite{GN} 
and in 
eight dimensions with gauge groups $SO(8)$ \cite{GKS}
and $SO(7)$ \cite{FN} have been found. 
Their construction started after it was realized that  
a higher-dimensional analog of the ordinary four-dimensional self-duality 
condition may be written down \cite{CDFN}. 
This condition is based on the octonionic 
algebra and leads to solutions similar to 
the standard four-dimensional 
Yang-Mills instantons. It also provides superstring solitonic 
configurations, the octonionic strings, 
 which preserve 1/16 of the space-time supersymmetries 
so that the world-sheet action  has $(0,1)$ supersymmetry \cite{HS}. 
In this case,  although the metric is asymptotically flat, the fields do 
not fall fast enough to give a finite ADM mass per unit length to the 
octonionic string. This divergence, however, is in the  infrared and 
it is of the type already encountered 
in the four-dimensional axionic string.   
Similarly,  supermembrane
solitonic solutions, the octonionic membranes, exist as well \cite{DM}. 
Here, the antiself-dual solution 
is asymptotically flat and has finite ADM mass while the self-dual one 
is not asymptotically flat. However, both solutions break all of the 
space-time supersymmetries.  

Recently, 
the octonionic self-duality condition has also been formulated for the 
supermembrane \cite{Fairly},\cite{F2}. It has been used  to 
exploit non-perturbative aspects of the vacuum by studying 
Nahm-type equations for the tunnelling processes between 
classical supermembrane configurations \cite{FL}. In seven dimensions, 
it has been shown that only one supersymmetry survives, while in three 
dimensions eight supersymmetries are preserved by the self-dual supermembrane
configuration \cite{FL2}.
It is also likely that octonionic 
monopole solutions to a Bogomolnyi equation should exist.  The octonionic 
algebra has also been employed in the compactifications of the 
eleven-dimensional supergravity on the round and squashed seven sphere
with or without torsion \cite{7S}.  
Moreover,  generalised self-duality equations for the 
p-forms have recently been discussed  \cite{BL}.

The close resemblance of Yang-Mills and gravitational instantons in 
four-dimensions leads to search for higher-dimensional gravitational 
instantons as well. Such a programme has not yet been carried out as far as 
we know, although some related work had already appeared \cite{AL}. 
Here, we will take first steps towards a systematic approach to these 
higher dimensional gravitational instantons. 
We will mainly study  eight-dimensional spaces 
with ``self-dual'' connection although 
a brief discussion for the seven-dimensional case will be given. 
We will see that manifolds which satisfy an appropriate self-duality 
condition in eight (seven) dimensions have  holonomy in $Spin(7)$ ($G_2$)
and this gives a systematic way of constructing such spaces by 
solving appropriate first-order equations.  
Non-compact manifolds of $Spin(7)$ holonomy have first been constructed in 
\cite{B} and compact ones as $T^8$ orbifolds in \cite{J7}
after appropriate resolving the singularities. Superstring  compactifications 
on $Spin(7)$ and $G_2$  manifolds have first been discussed in \cite{VS}
where the corresponding superconformal algebras were given and latter in 
\cite{FFO}. Here, 
by solving the eight-dimensional   self-duality 
condition we find a space with holonomy in $Spin(7)$. 
It has an elliptically-fibered structure where the base is 
a sphere and the fibres are three tori which degenerate at 
twenty-four points. Its four-dimensional counterpart is the stringy cosmic
string \cite{CS}. 

In the following section, we will recall some results
concerning the octonionic algebra and its relation to $SO(8)$. In
section 3, we will discuss eight-dimensional 
self-dual manifolds  and in section
4, we will present a specific solution to the self-duality 
condition. Finally, in section 5, we  extend our results 
to seven dimensions.

\section{Octonionic algebra}

We will recall here some properties of the octonionic 
(Cayley) algebra ${\cal{O}}$ and its relation to $SO(8)$
which we will use  in the sequel. We follow mainly \cite{GG},\cite{dN}. 
The octonionic algebra is a division 
algebra, which means it has a non-degenerate quadratic form $Q$ that satisfies
$Q(xy)=Q(x)Q(y)$ and in addition $Q(x)=0$  implies $x=0$. The other division 
algebras are the real ${\cal R}$, complex ${\cal C}$ and quaternionic 
algebras. ${\cal R}$, ${\cal C}$ are commutative, 
${\cal H}$ is non-commutative 
while ${\cal {O}}$ is neither commutative nor associative.   
A basis for ${\cal{O}}$ is provided by the eight elements
\be
1,~e_a, ~~~~~ a=1,\ldots,7, 
\ee
which satisfy the relation 
\be
e_ae_b=\psi_{abc}e_c-\delta_{ab}\, .
\ee
It is similar to the corresponding relation obeyed by the quaternions
(Pauli matrices). However, the latter satisfy the Jacobi identity as a 
result of the associativity of the quaternionic algebra, 
while the octonions $e_a$ are not associative
and do not satisfy the Jacobi identity.   
The tensor $\psi_{abc}$ is totally antisymmetric with 
\be
\psi_{abc}=+1~~~{\mbox{for}}~~~~abc=123,~516,~624,~435,~471~,673,~572.
\ee
We may also define its dual $\psi_{abcd}$ as
\be
\psi_{abcd}=\frac{1}{3!} \epsilon_{abcdfgh}\psi^{fgh}\, , \label{psi}
\ee
so that 
\be
\psi_{abcd}=+1~~~ {\mbox{for}}~~~~abcd=1245,~2671,~3526,~4273,~5764,~6431,~
7531.
\ee 
They satisfy the relations 
\be
\psi^{abc}\psi_{dhc}&=&\d^a_d\d^b_h-\d^a_h\d^b_d-{\psi^{ab}}_{dh}\, , 
\nonumber\\
\psi^{abcd}\psi_{ehcd}&=&4\left(\d^a_e\d^b_h-\d^a_h\d^b_e\right)-
2{\psi^{ab}}_{eh}\, , \nonumber \\
\psi^{abc}\psi_{debc}&=&-4{\psi^{a}}_{de}\, . 
\ee
The tensor $\psi_{abc}$ can be assigned to an $SO(8)$ representation 
$\Psi_{\a\b\g\d}$, $\a,\ldots=1,\ldots,8$ as
\begin{eqnarray}
\Psi_{\a\b\g 8}=\psi_{abc}\, , ~~~~
\Psi_{\a\b\g\d}=\psi_{abcd}\, , ~~~~~~\a=(a,8)\,\ldots \, , \label{p}
\end{eqnarray}
which is self-dual 
\be
\Psi_{\a\b\g\d}=\frac{1}{4!}\epsilon_{\a\b\g\d\zeta\eta\theta\kappa}
\Psi^{\zeta\eta\theta\kappa}\, , \label{pe}
\ee
and belongs to one of the three different ${\bf 35}$'s of $SO(8)$ ${\bf 35}_v$,
${\bf 35}_\pm$ 
(related by triality). 
It satisfies the fundamental identity
\be
\Psi_{\a\b\g\d}\Psi^{\zeta\eta\theta\d}&=&
\big{(}\d_\a^\zeta\d_\b^\eta-\d_\b^\zeta\d_\a^\eta)\d_\g^\theta+
(\d_\a^\theta\d_\b^\zeta-\d_\b^\zeta\d_\a^\theta)\d_\g^\eta+
(\d_\a^\eta\d_\b^\theta-\d_\b^\theta\d_\a^\eta)\d_\g^\eta+
\nonumber \\
&&{\Psi_{\a\b}}^{\zeta\eta}\delta_\g^\theta
+{\Psi_{\a\b}}^{\theta\zeta}\delta_\g^\eta+
{\Psi_{\a\b}}^{\eta\theta}\delta_\g^\zeta +
{\Psi_{\g\a}}^{\zeta\eta}\delta_\b^\theta+
{\Psi_{\g\a}}^{\theta\zeta}\delta_\b^\eta+
\nonumber \\&&{\Psi_{\g\a}}^{\eta\theta}\delta_\b^\zeta+
{\Psi_{\b\g}}^{\zeta\eta}\delta_\a^\theta+
{\Psi_{\b\g}}^{\theta\zeta}\delta_\a^\eta+
{\Psi_{\b\g}}^{\eta\theta}\delta_\a^\zeta \, . \label{fund}
\ee 
By contraction of the above identity we get
\be
\Psi_{\a\b\g\d}\Psi^{\zeta\eta\g\d}&=&6(\d_\a^\zeta\d_\b^\eta-\d_\b^\zeta
\d_a^\eta)+4{\Psi_{\a\b}}^{\zeta\eta}\, , \label{pp}
%\\ \Psi_{\a\b\g\d}\Psi^{\zeta\b\g\d}&=&42\d_\a^\zeta \, , \label{ppp}\\
%\Psi_{\a\b\g\d}\Psi^{\a\b\g\d}&=&336\, . \label{pppp}
\ee

We may use the octonions $e_a$  to construct a 
 representation of the $SO(7)$ $\g$-matrices according to 
\be
\left(\gamma_a\right)_{bc}=i\psi_{abc}\, , ~~~\left(\gamma_a\right)_{b8}=i
\delta_{ab}\, , ~~~~~\{\gamma_a,\gamma_b\}=2\delta_{ab}\, ,
\ee
so that $\g^{ab}=\frac{1}{2}[\g^a,\g^b]$ are the $SO(7)$ generators. 
We may also form the $SO(8)$ $\g$-matrices $\Gamma_\a=(\Gamma_a,\Gamma_8)\,, 
\a=1,\ldots,8$
\be
\Gamma_a=\left(\matrix{0&i\gamma_a\cr -i\gamma_a&0}\right)\, , 
~~~~~~ \Gamma_8=
\left(\matrix{0&1\cr 1&0}\right)\, , ~~~~~~~
\{\Gamma_\a,\Gamma_\b\}=2\d_{\a\b}\, , 
\ee
that correspond to the standard embedding 
of $SO(7)_v$ in $SO(8)$. 
The latter is defined as the stability subgroup  $SO(7)\subset 
SO(8)$ of the vector 
representation according to which we have the decomposition
\be
{\bf 8}_v={\bf 7}+{\bf 1}\, , ~~~{\bf 8}_{\pm}={\bf 8}\, , 
\ee
where ${\bf 8}_v,{\bf 8}_{\pm}$ are the vector and the two spinorial 
representations of $SO(8)$.  
The $SO(8)$ generators $\Gamma^{ab}=\frac{1}{2}[\Gamma^a,\Gamma^b]$ 
satisfy the relations 
\be
\Gamma^{ab}=\psi^{abc}\Gamma^9\Gamma_{8c}\, ,~~~
\psi_{abcd}\Gamma^{ab}=-(4+2\Gamma^9)\Gamma_{cd}\, , 
 \label{8c}
\ee
where, as usual, $\Gamma^9=\left(\matrix{1&0\cr 0&-1}\right)$ is the chirality 
matrix. 
It is not then difficult to verify that the generators
\be
G^{\a\b}=\frac{3}{8}\left(\Gamma^{\a\b}+\frac{1}{6}{\Psi^{\a\b}}_{\g\d}
\Gamma^{\g\d}\right)\, , \label{G} 
\ee
leave the right-handed spinor $\eta_+$ invariant
\be 
G^{\a\b}\eta_+=0 
%~~~~ \eta_\a=\delta_{8\a}
\, . \label{Gh}
\ee
The stability 
group of $\eta$ is again another $SO(7)$, which we will denote by 
$SO(7)^+$,
according to which  
\be
{\bf 8}_v={\bf 8}\, , ~~{\bf 8}_+={\bf 7}+{\bf 1}\, , 
~~{\bf 8}_-={\bf 8}
\, . \label{8+}
\ee
The singlet in the decomposition of ${\bf 8}_+$ corresponds to the null
eigenspinor $\eta$ of $G^{\a\b}$ in eq.(\ref{Gh}). This can be seen by 
considering the Casimir $G^{\a\b}G_{\a\b}$ which has a zero eigenvalue 
for the ${\bf 8}_+$. 
If we had used in eq.(\ref{G}) the second ${\bf 35}$ of $SO(8)$, the 
antiself-dual one, then $G^{\a\b}$ would annihilate the left-handed 
spinor $\eta_-$. In this case, the stability group of the latter 
is a  third $SO(7)^-$ subgroup of $SO(8)$ defined by
\be
{\bf 8}_v={\bf 8}\, ,  ~~{\bf 8}_+={\bf 8}\,,~~{\bf 8}_-={\bf 7}+{\bf 1}\, .
\ee
Again, $G^{\a\b}$ constructed with the antiself-dual ${\bf 35}$ has a 
zero eigenvalue which corresponds to the singlet in ${\bf 8}_-$. 
The three different $SO(7)$ subgroups of $SO(8)$, $SO(7)_v,
SO(7)^\pm$ are related by triality.

%\section{Some properties of $Spin(7)$ manifolds}

%The bosonic part of the eleven-dimensional supergravity is, in an obvious 
%notation
%\be
%S=\frac{1}{2\kappa_{11}^2}
%\int d^{11}x \left(R-\frac{1}{2\cdot9!}
%F_{M_1\ldots M_4}F^{M_1\ldots M_4}+\frac{2}{(72)^2}
%\epsilon^{M_1\ldots M_{11}}F_{M_1\ldots M_4}F_{M_5\ldots M_8}
%A_{M_9M_{10}M_{11}}\right)
%\ee
%where $F_{M_1M_2M_3M_4}=\partial_{M_1}A_{M_2M_3M_4}+\!cyclic$
%is the four-form field strength and $\kappa_{11}$ the $D=11$ gravitational 
%coupling constant. By dimensional reduction of the above action we get 
%the various supergravity theories. Here, we will discuss vacuum configurations
%in eight dimensions which preserve some supersymmetry. 

\section{$Spin(7)$ holonomy spaces}

By dimensional reduction of the eleven-dimensional supergravity, other 
supergravity theories are obtained. For example, reduction of the former 
on a three torus results in the $N=2$ eight-dimensional supergravity.  
The graviton multiplet contains the graviton, an antisymmetric three-form,
three two-forms, six one-forms and six scalars and a
generic vacuum configuration  breaks  all supersymmetries. 
As usual, when all fields vanish except graviton, the condition for unbroken 
supersymmetry in a pure gravitational background  is that 
the gravitino shifts are zero, i.e., 
\be
\delta\psi_\mu^A=\left(\partial_\mu-\frac{1}{4}\omega_{\alpha\beta\mu}
\Gamma^{\alpha\beta}\right)\eta^A
=D_\mu\eta^A=0\, , \label{eta}
\ee
where $\alpha,\beta,...=1,...,8$ are world indices, $\mu,\nu,...=1,...,8$
curved space ones, $A=1,2$ counts the two supersymmetries of opposite chirality
and  $\omega_{\alpha\beta\mu}$
is the spin connection. 
The integrability condition of eq.(\ref{eta}) is 
\be
R_{\a\b\mu\nu}\Gamma^{\a\b}\eta^A=0\, , 
\ee
from which it follows, after multiplying by $\Gamma^\mu$ that such a
background is Ricci flat. 
The obvious solution of eq.(\ref{eta}) is flat space.
We will see that there also exist  non-trivial solutions to eq.(\ref{eta})
beyond the trivial flat space. 

Let us suppose that the spin connection satisfies the 
``self-duality'' condition
\be
 \omega_{\alpha\beta\mu}=\frac{1}{2}\Psi_{\a\b\g\d} 
{\omega^{\g\d}}_{\mu}
\, , \label{sd}
\ee
where $\Psi_{\a\b\g\d}$ is defined in eq.(\ref{p})
\footnote{A relation of the form  
$\omega_{\alpha\beta\mu}= \lambda\Psi_{\a\b\g\d} 
{\omega^{\g\d}}_{\mu}$ is consistent for $\lambda=1/2,-1/6$ as can be 
verified by multiplying both sides 
by $\Psi^{\a\b\zeta\eta}$ and using eq.(\ref{pp}).
Here we will consider only the case $\lambda=1/2$.}.
It is not then difficult to verify that the spinor $\eta^A$ in 
eq.(\ref{eta}) is necessarily right-handed and 
\be
\d\psi_\mu= 
\partial_\mu\eta^A -\frac{1}{2}\omega_{\a\b\mu}G^{\a\b}\eta^A\, ,
%\Gamma^{ab}\eta^A -\frac{1}{2}\omega_{a8\mu}
%\Gamma^{a8}\eta^A=\partial_\mu\eta^A -\frac{1}{4}\omega_{ab\mu}
%\left(\Gamma^{ab}\pm \Psi^{abc}\Gamma^9\Gamma_{8c}\right) \eta^A
%\, , 
\ee
where $G^{\a\b}$ has been defined in eq.(\ref{G}).  
We see that 
the gravitino shifts vanish if $\eta^A$ satisfies eq.(\ref{Gh})
which means that the background preserves 
$1/16$ of the original supersymmetries. 
The Killing spinor is invariant under $SO(7)^+$ and thus, 
spaces with connection satisfying eq.(\ref{sd}) have holonomy in 
$Spin(7)$. 

Eq.(\ref{sd}) is an eight-dimensional analog of the standard
self-duality condition in four dimensions  
where the quantity $\Psi_{\a\b\g\d}$  replaces 
by the totally antisymmetric symbol $\epsilon_{abcd}$. 
The corresponding solutions are the four-dimensional gravitational
instantons. In this respective, we will also call eight-dimensional 
manifolds with connection satisfying  
eq.(\ref{sd}) in the Euclidean regime gravitational instantons. 
As we have already pointed out, the holonomy group 
of such manifolds is in $Spin(7)$. Alternatively, this can be
seen as follows. Let us recall that a manifold is of $Spin(7)$ holonomy 
if  and only if the 
Cayley four-form \cite{sal} 
\be 
\Psi&=&e^1\wedge e^2\wedge e^3\wedge e^8+ e^5\wedge e^1\wedge e^6\wedge e^8+
e^6\wedge e^2\wedge e^4\wedge e^8+e^4\wedge e^3\wedge e^5\wedge e^8\nonumber \\
&&+e^4\wedge e^7\wedge e^1\wedge e^8+e^6\wedge e^7\wedge e^3\wedge e^8
+e^5\wedge e^7\wedge e^2\wedge e^8+e^4\wedge e^5\wedge e^6\wedge e^7\nonumber\\
&&+e^2\wedge e^3\wedge e^7\wedge e^4+e^1\wedge e^3\wedge e^5\wedge e^7
+e^1\wedge e^3\wedge e^5\wedge e^7+e^1\wedge e^2\wedge e^7\wedge e^6 
\nonumber \\
&&+e^2\wedge e^3\wedge e^5\wedge e^6
+e^1\wedge e^2\wedge e^4\wedge e^5
+e^1\wedge e^3\wedge e^4\wedge e^6\, , \label{c}
\ee
where $e^\a,~ \a=1,\ldots,8$ is an orthonormal frame (the metric is 
$\sum_ie^i\otimes e^i$ in this frame), is torsion free, i.e., if it is closed 
\be
d\Psi=0\, . 
\ee
In this case, the manifold is Ricci flat.
The Cayley form  $\Psi$ is self-dual and, 
in addition, it is invariant under $Spin(7)$. It is the singlet in the 
decomposition of the ${\bf 35}_+$ in  the non-standard embedding  
of $SO(7)$  in $SO(8)$ given in 
eq.(\ref{8+})  according to which  
\be
{\bf 35}_v={\bf 35}\, , ~~~  {\bf 35}_+={\bf 1} +
{\bf 7}+{\bf 27}\, ,  ~~~ {\bf 35}_-={\bf 35}\, .
\ee

In order to prove now 
that manifolds with connection that satisfies eq.(\ref{sd})
are of $Spin(7)$ holonomy, let us observe that the Cayley four-form $\Psi$ 
can actually be written as 
\be
\Psi=\frac{1}{4!}\Psi_{\a\b\g\d}e^\a\wedge e^\b\wedge e^\g\wedge e^\d\, . 
\label{omega}
\ee 
It is not difficult then to verify by using eq.(\ref{fund}) and the structure 
equations 
$$de^\a+{\omega^\a}_\b e^\b=0\, , ~~~~~~{\omega^\a}_\b=
{\omega^\a}_{\b\mu} dx^\mu\, , $$
 that, if the spin connection satisfies 
eq.(\ref{sd}), the Cayley form $\Psi$ is indeed closed. Thus, manifolds 
whose connection satisfies the ``self-duality'' condition eq.(\ref{sd}) 
are  Ricci flat $Spin(7)$ manifolds.

\section{Eight-dimensional gravitational instantons}

Our aim is to construct manifolds of $Spin(7)$ holonomy by solving
 eq.(\ref{sd}). By choosing  one direction, the eighth say,  
eq.(\ref{sd}) is written as 
\be
{\omega^8}_r=-\frac{1}{2}\psi_{rpq}\omega^{pq}\, . \label{seq}
\ee
The rest of the equations, namely,
\be
{\omega^p}_q=\frac{1}{2}{\psi^p}_{qrs}\omega^{rs}-
{\psi^p}_{qr}\omega^{r8}\, , \label{seq2} 
\ee
are automatically satisfied if eq.(\ref{seq}) holds. The self-duality 
conditions are then explicitly written as 
\be
\omega_{81}=-\left(\omega_{23}+\omega_{65}+\omega_{47}\right)\, , &&
\omega_{82}=-\left(\omega_{31}+\omega_{46}+\omega_{57}\right)\, ,\nonumber\\ 
\omega_{83}=-\left(\omega_{12}+\omega_{54}+\omega_{67}\right)\, , &&
\omega_{84}=-\left(\omega_{62}+\omega_{35}+\omega_{71}\right)\, ,\nonumber\\
\omega_{85}=-\left(\omega_{16}+\omega_{43}+\omega_{72}\right)\, , &&
\omega_{86}=-\left(\omega_{51}+\omega_{24}+\omega_{73}\right)\, ,\nonumber\\
\omega_{87}=-\left(\omega_{14}+\omega_{36}+\omega_{25}\right)\, ,  &&   
\label{expli}
\ee
The obvious solutions to the above equations 
are spaces of the form $M_1\times M_2$ where $M_1,M_2$
are four-dimensional manifolds of self-dual or antiself-dual connections. 
However, other solutions exist as well. Here we will present a space 
with  metric
\be
ds^2&=&R^2(dx_1^2+dx_2^2)+V^2\left(dx_3+2Adx_8\right)^2+U^2dx_8^2
+\label{metric}\\ &&
E^2\left(dx_4+2Bdx_5\right)^2+
G^2dx_5^2+H^2dx_6^2+F^2\left(dx_7+2Cdx_6\right)^2\, , \nonumber
\ee
where $R,V,A,U,E,B,G,F,C,H$ are functions of $x_1,x_2$ only and they will be 
specified by solving eqs.(\ref{expli}). We may introduce the 
orthonormal basis
\be
&&e^1=Rdx_1\, ,~~e^2=Rdx_2\, ,~~e^3=V(dx_3+2Adx_8)\, ,~~
e^4=E(dx_4+2Bdx_5)\, , \nonumber
\\
&&e^5=Gdx_5\,,
~~e^6=Hdx_6\,, ~~e^7=F(dx_7+2Cdx_6)\, ,~~ e^8=Udx_8\,, \label{basis} 
\ee
in which the metric is $ds^2=\sum(e^i)^2$. In this base,  
the non-zero components of the connection are
\be
&&\omega_{12}=\frac{1}{R^2}\p_2Re^1-\frac{1}{R^2}\p_1Re^2\, , ~~
\omega_{31}=\frac{V}{RU}\p_1Ae^8+\frac{1}{RV}\p_1Ve^3\,,\nonumber \\&&
\omega_{32}=\frac{V}{RU}\p_2Ae^8+\frac{1}{RV}\p_2Ve^3\,,~~
\omega_{41}=\frac{E}{RG}\p_1Be^5+\frac{1}{RE}\p_1Ee^4\,,\nonumber \\&&
\omega_{42}=\frac{E}{RG}\p_2Be^5+\frac{1}{RE}\p_2Ee^4\,,~~
\omega_{51}=\frac{1}{RG}\p_1Ge^5+\frac{E}{RG}\p_1Be^4\,,\nonumber \\&&
\omega_{52}=\frac{1}{RG}\p_2Ge^5+\frac{E}{RG}\p_2Be^4\,,~~ 
\omega_{61}=\frac{1}{RH}\p_1He^6+\frac{F}{RH}\p_1CFe^7\,, \nonumber \\&&
\omega_{62}=\frac{1}{RH}\p_2He^6+\frac{F}{RH}\p_2CFe^7\,,~~
\omega_{71}=\frac{F}{RH}\p_1Ce^6+\frac{1}{RF}\p_1Fe^7\,,\nonumber \\&&
\omega_{72}=\frac{F}{RH}\p_2Ce^6+\frac{1}{RF}\p_2Fe^7\,,~~
\omega_{81}=
\frac{1}{RU}\p_1Ue^8+\frac{V}{RU}\p_1Ae^3\,,\nonumber  \\&&
\omega_{82}=\frac{1}{RU}\p_2Ue^8+\frac{V}{RU}\p_2Ae^3\,,~~
\omega_{83}=\frac{V}{RU}\p_1Ae^1+\frac{V}{RU}\p_2Ae^2\, , \nonumber \\&&
\omega_{54}=\frac{E}{RG}\p_1Be^1+\frac{E}{RG}\p_2Be^2\, ,~~ 
\omega_{67}=\frac{F}{RH}\p_1Ce^1+\frac{F}{RH}\p_2Ce^2\, . \label{conn}
\ee
where $\p_1=\p/\p x^1,\p_2=\p/\p x^2$. 
%$$R=R(x_1,x_2),V=V(x_1,x_2),A=A(x_1,x_2), U=U(x_1,x_2), E=E(x_1,x_2),
%B=B(x_1,x_2), G=G(x_1,x_2),F=F(x_1,x_2),C=C(x_1,x_2),H=H(x_1,x_2)$$.
Substituting the above expressions in eq.(\ref{expli}), we get 
that 
\be
V=\frac{1}{U}\, , ~~E=\frac{1}{G}\, , ~~ F=\frac{1}{H}\, , ~~R=UHG\, .
\ee
In addition, the functions $2A,U^2,2B,G^2,2C,H^2$ satisfy 
the Cauchy-Riemann equations
\be
&&\p_1U^2=2\p_2A\, , ~~\p_1G^2=2\p_2B\, , ~~\p_1G^2=2\p_2B\, , \nonumber \\&&
\p_2U^2=-2\p_1A\, , ~~\p_2G^2=-2\p_1B\, ,~~ \p_2G^2=-2\p_1B\, ,\label{CR}
\ee
and the metric is then  written as 
\be
ds^2&=&U^2G^2H^2(dx_1^2+dx_2^2)+U^{-2}\left(dx_3+2Adx_8\right)^2+U^2dx_8^2+
\label{metric1}\\ && G^{-2}\left(dx_4+2Bdx_5\right)^2+
G^2dx_5^2+H^2dx_6^2+H^{-2}\left(dx_7+2Cdx_6\right)^2\, . \nonumber
\ee
We may now introduce complex functions 
\be
\tau=2A+iU^2\, , ~~\sigma=2B+iG^2\, , ~~\rho=2C+iH^2\, , \label{complex}
\ee
so that eqs.(\ref{CR}) are satisfied for holomorphic $\tau=\tau_1+i\tau_2,
\sigma=\sigma_1+i\sigma_2,\rho=\rho_1+i\rho_2$,
\be
\tau=\tau(z)\, ,~~\sigma=\sigma(z)\, , ~~\rho=\rho(z)\, ,
\ee
where $z=x_1+ix_2$. Then, the metric turns out to be
\be
ds^2&=&\tau_2\sigma_2\rho_2|h(z)|^2dzd\bar{z}+\frac{1}{\tau_2}
|dx_3+\tau(z)dx_8|^2 +\nonumber \\ &&\frac{1}{\sigma_2}
|dx_4+\sigma(z)dx_5|^2 +\frac{1}{\rho_2}
|dx_7+\rho(z)dx_6|^2 \, , \label{metrict}
\ee
where $h(z)$ is an arbitrary holomorphic function of $z$. From 
eq.(\ref{complex}) we see that $\tau,\sigma,\rho$ belong to the upper half 
plane and thus each parametrize an $SL(2,{\bf R})/U(1)$ coset space. In
fact, the metric is 
invariant under $SL(2,{\bf R})\!\times\!SL(2,{\bf R})\!\times\!SL(2,{\bf R})$ 
which acts as 
\be
&&\tau\to \frac{a\tau+b}{c\tau+d}\,, ~~~
x_3\to -bx_8+ax_3\, , ~~~x_8\to dx_8-cx_3\, ~~~~ ad-bc=1\, , \nonumber \\
&& \sigma\to \frac{k\sigma+\ell}{m\sigma+n}\,, ~~~ x_4\to-\ell x_5+kx_4\, ,~~~
x_5\to nx_5-mx_4\, ,~~~~ kn-\ell m=1\, , \nonumber 
 \\
&& \rho\to \frac{p\rho+r}{s\rho+q}\,, ~~~
x_6\to qx_6-sx_7\, , ~~~ x_7\to-rx_6+px_7\,,~~~~ pq-rs=1\, . 
\ee
For periodic $x_3,\ldots,x_8$, the transverse space to the z-plane is a 
product of three $T^2$. In that case,
we may view $\tau,\sigma,\rho$ as the modulus of the tori if we restrict 
them to the fundamental domain of $SL(2,{\bf Z})$. Under the latter 
they transform as 
\be
\tau\to \tau+1\, , ~~\tau\to-\frac{1}{\tau}\, , ~~~
\sigma \to \sigma +1\, , ~~\sigma\to-\frac{1}{\sigma}\, ,~~~
\rho\to \rho+1\, , ~~\rho\to-\frac{1}{\rho}\, .
\ee  
The holomorphic functions $\tau(z),\sigma(z),\rho(z)$ will have
certain singularities in the z-plane which can be interpreted as magnetic 
sources. In that case, as we go around these singularities we will have the 
$SL(2,{\bf Z})$ jumps $\tau\to \tau+1,\sigma\to \sigma+1,\rho\to \rho+1$
\cite{CS},\cite{K} so that, 
 if the singularities of $\tau,\sigma,\rho$ are at 
$z_i^\tau,z_i^\sigma,z_i^\rho$, respectively,
\be
\tau\sim -\frac{i}{2\pi}\log (z-z_i^\tau)\, ,~~
\sigma\sim -\frac{i}{2\pi}\log (z-z_i^\sigma)\, ,
~~\rho\sim -\frac{i}{2\pi}\log (z-z_i^\rho)\,.
\ee
The metric (\ref{metrict}) is $SL(2,{\bf Z})$ invariant
if $h(z)$ in eq.(\ref{metrict}) has certain modularity properties 
without zeroes in the fundamental domains. This uniquely specifies 
\be
h(z)=\eta(\tau)^2\eta(\sigma)^2\eta(\rho)^2\, , \label{h}
%\prod_{i,j,k=1}^{N_\tau,N_\sigma,N_\rho}
%\prod_{j=1}^{N_\sigma}\, ,  
\ee
up to a modular invariant holomorphic function, 
where $$\eta(\tau)=q^{1/24}\prod_n(1-q^n)\, , $$
is the Dedekind's function ($q=exp(2\pi i\tau)$). 
However, since $q\sim 
(z-z_i)$, we have that 
$$\eta(\tau)^2\eta(\sigma)^2\eta(\rho)^2\sim
\left((z-z_i^\tau)(z-z_j^\sigma)(z-z_k^\rho)\right)^{1/12}$$
and thus instead of eq.(\ref{h}) we should take  
\be
h(z)=\eta(\tau)^2\eta(\sigma)^2\eta(\rho)^2
\prod_{i,j,k=1}^{N_\tau,N_\sigma,N_\rho}
%\prod_{k=1}^{N_\rho}
\left(z-z_i^\tau\right)^{-1/12}\left(z-z_j^\sigma\right)^{-1/12}
\left(z-z_k^\rho\right)^{-1/12}\,,
\ee
where $N_\tau,N_\sigma,N_\rho$ are the number of singularities of 
$\tau,\sigma,\rho$, respectively. As usual, we must have 
$N_\tau+N_\sigma+N_\rho=12$ 
in order that no conical singularities appear. In this case, the z-plane is 
asymptotically a cylinder and if $N_\tau+N_\sigma+N_\rho=24$, we have 
a sphere \cite{CS}. 

\section{Comments}
We have discussed here eight-dimensional self-dual spaces. As we have seen, 
such spaces have holonomy in $Spin(7)$. In a sense, they generalize 
the four-dimensional self-dual manifolds which are of $SU(2)$ holonomy. 
We have presented a space with a self-dual connection and an 
elliptically-fibered structure. 
The base is a sphere and the fibres are three tori. The tori
degenerate at a total number of twenty-four  points in the same way as in 
the stringy cosmic string solution \cite{CS},\cite{K}. 

It should be noted that a self-duality condition can also be written down in 
seven dimensions as
\be
\omega_{ab}=\lambda\psi_{abcd}\omega^{cd}\, , \label{g2}
\ee
where $\psi_{abcd}$ have been defined in eq.(\ref{psi}) and 
$\lambda=1/2,-1/4$. 
In this case,  spaces with connection satisfying the 
above equation are of holonomy in $G_2$. This can be seen by recalling
that for the $G_2$ case, the Cayley four-form is replaced by a three 
form $\Phi$ and its dual $\ast\Phi$ which turn out to be
\be
\Phi=\frac{1}{3!}\psi_{abc}e^a\wedge e^b\wedge e^c\, , ~~
\ast \Phi =\frac{1}{4!}\psi_{abcd} e^a\wedge e^b\wedge e^c\wedge e^d\, .
\ee
The condition of $G_2$ holonomy then reduces to the closure of $\Phi,\ast\Phi$.
Indeed, it is not difficult to prove that for spaces with connection 
satisfying eq.(\ref{g2}), $\Phi$ and $\ast\Phi$ are closed.  

Spaces of $G_2$ holonomy are  relevant for M-theory compactifications. 
 They give rise to N=1 in four-dimensions and they are the counterparts of 
CY compactifications of string-theory. As have been pointed out in 
\cite{J2}, $G_2$ holonomy spaces can be constructed from CY threefolds  
as the quotient $(CY_3\times S^1)/\pi$ where $\pi$ acts as a complex
involution on $CY_3$ 
and as $x\to-x$ on $S^1$ and then appropriately resolving
the singularities. Non-compact examples have been constructed in 
\cite{B}. 

An interesting open problem, mentioned also in \cite{J7}, is the construction 
of eight-dimensional ALE spaces. For example, there should exists a  
 higher-dimensional analog 
of the Eguchi-Hanson solution  which is the simplest case of 
the four-dimensional ALE spaces with $SU(2)$ holonomy. Such manifolds, 
as well as ALE spaces of $SU(n)$ holonomy, 
have extensively been studied. However, nothing is actually  known about 
ALE spaces of $Spin(7)$ of $G_2$ holonomy. We believe, that there should exist
solutions to the self-duality condition eq.(\ref{sd}) which will correspond
to such spaces.

%\noindent
%{\bf{Acknowledgement}} 
%\vspace{.3cm}

%\newpage

%\begin{flushleft}
%{\large\bf Appendix}\end{flushleft}
%\renewcommand{\theequation}{A.\arabic{equation}}
%\renewcommand{\thesection}{B.}
%\setcounter{equation}{0}

\end{document}